\documentstyle[12pt]{article}
\def\be{\begin{eqnarray}}
\def\ee{\end{eqnarray}}
\def\ba{\begin{array}}
\def\ea{\end{array}}

\begin{document}

\begin{center}
{\bf\Large {Gravity as Classical Effect in Generalized Relativistic
\\\vskip 3mm Quantum Mechanics on Flat Background}}
\end{center}

\vskip 1cm

\begin{center}
{\bf \large {Vadim V.
Asadov$^{\,\star\,}$\footnote{asadov@neurok.ru} and Oleg V.
Kechkin$^{\,+\,\star\,}$}\footnote{kechkin@depni.sinp.msu.ru}}
\end{center}

\vskip 1cm

\begin{center}
$^+\,$Institute of Nuclear Physics,\\
Lomonosov Moscow State University, \\
Vorob'jovy Gory, 119899 Moscow, Russia
\end{center}

\vskip 3mm

\begin{center}
$^\star\,$Neur\,OK-III,\\
Scientific park of MSU, Center for Informational Technologies--104,\\
Vorob'jovy Gory, 119899 Moscow, Russia
\end{center}

\vskip 1cm

\begin{abstract}
We study classical limit for quantum mechanics with two times and
temperature, which describes a generalized dynamics of relativistic
point mass. In this theory, thermodynamic time means a parameter of
evolution, whereas geometric time is one of space-time coordinates.
We identify world lines in this theory with geodesic lines for the
same point mass in standard General Relativity in its weak gravity
limit. In identification performed, effective metrics is generated
by weak correlations between canonical variables of the originally
flat relativistic mechanics, i.e. between its space-time coordinates
and moments.
\end{abstract}

\vskip 0.5cm

PACS No(s).\, : 05.30.-d,\,\,05.70.-a.

\vskip 1cm

Unification of quantum theory with thermodynamics and geometry is
one of the most important problems in the modern theoretical
physics. The unified theoretical scheme must contain all fundamental
and the most general conceptions of these three `supertheories';\,
its the most natural realizations -- quantum black hole physics and
cosmology -- will satisfy the second low of thermodynamics up to
construction. Such introducing of arrow of time
(\cite{atf}--\cite{atl}) into the structure of theory will clarify
dramatically our understanding of evolution and present structure of
the total Universe and its non-trivial geometric constituents. A
construction of any consistent quantum gravity model with arrow of
time will be really important input into realization of the general
programm formulated above.

In \cite{q-1} it was shown, that unification of quantum and
thermodynamic principles can be performed on the base of the first
ones. Namely, the unidirectional evolution of dynamical system takes
place in the quantum theory with complex time parameter and
non-Hermitian Hamiltonian structure. This theory is postulated as
analytical in respect to the parameter of evolution, which real part
is identified with the `usual'\, physical time, whereas the
imaginary one is understood as proportional to the inverse absolute
temperature. Also, the Hermitian part of the Hamiltonian is put
equal to conventional operator of energy. It is shown, that the
anti-Hermitian Hamiltonian part, which is taken as commuting with
the energy operator, is constructed from parameters of decay of the
system. It is established, that quantum dynamics, predicted by this
theory, possesses a well defined arrow of time in isothermal and
adiabatic regimes of the evolution. In \cite{q-2} it was also
established, that the presence of arrow of time in the quantum
dynamics can be a result of parity violation in the system. It takes
place if the decay operator is taken as proportional to the parity
one, and for extremely wide class of thermodynamic regimes of the
evolution. Finally, in \cite{q-c} it was shown, that classical limit
of the quantum theory of the discussing type is also
time-irreversible. The main idea of these works is related to fact,
that it is possible `to derive'\, the thermodynamics from the
quantum theory modified in some really `minimal sense'.\, In this
paper we would like to show, that the same generalized quantum
theoretical framework contains a gravity as its pure classical
limit. It seems, that one needs in consequent `putting of the
standard quantum principles to work'\, only to reach the great
unification scheme, discussed at the beginning of this paper.

Of course, in the gravitational context, it is necessary to start
with some Lorenz-invariant model at least to have any chance `to
derive gravity'\, from the quantum principles. In our previous work
\cite{r-1}, we have developed both quantum and classical variants
for relativistic theory of a point mass, which is based on the use
of two different types of time -- the thermodynamic and geometric
ones. The first time plays a role of parameter of evolution, whereas
a presence of the second one makes the theory Lorenz-invariant in
explicit form. It was demonstrated, that these independent time
parameters become proportional in a `hard'\, classical variant of
the theory, when one puts not only $\hbar=0$ in all originally
quantum relations, but also restricts a resulting probability
density by its delta-distribution realization. In fact, the
thermodynamic time is equivalent to self-time of the particle in the
hard dynamics mentioned.

Then, a `soft'\, (or the most general) variant of the modified
classical dynamics developed contains the non-trivial probability
density as one of the main elements of the formalism. This leads to
incorporating of coordinate-momentum correlations into equations of
motion of the theory. In this work we show, that corresponding
non-localized (in the probability density sense) classical particle
moves in the flat space-time like its standard analogy (which is
localized completely) in the curved space-time of the General
Relativity. We make the comparison and calculate an effective
space-time metrics in terms of the probability density related
quantities and other kinematical elements of the generalized
classical mechanics under consideration. Our main goal is to
demonstrate a possibility to interpret the standard classical
gravity as some correlation related classical effects of the
corresponding quantum relativistic theory with flat background.

We start with the statement, that in the `soft'\, classical variant
of the relativistic mechanics under discussion, one deals with pair
of the generating functions $S_a$ and $S_e$, which have the
action-like and entropy-like sense in the original quantum theory
framework \cite{r-1}. These functions depend on the thermodynamical
time $t$, and on the set of the space-time coordinates $x^{\mu}$
(where $\mu=0,...,3$ in the case of four-dimensional theory). Note,
that we do not suppose any relation between  $t$ and the geometric
coordinate $x^{0}$, because these variables are understood as the
completely independent ones on the kinematical level.

Then, the equations on the generating functions read:
\be\label{rel1-1}
S_{a,\,t}&=&\frac{1}{m}g_0^{\mu\nu}S_{a,\,\mu}S_{a,\,\nu},\\\label{rel1-2}S_{e,\,t}&-&\frac{2}{m}g_0^{\mu\nu}
S_{a,\,\mu}S_{e,\,\nu}=-\frac{2}{m}\Box S_{a},\ee where
$\Box=g_{(0)}^{\mu\nu}\partial_{\mu}\partial_{\nu}$, and
$g_{(0)}^{\mu\nu}={\rm diag} (-\,+\,+\,+)$ is the flat Minkowsky
metrics. These equations must be completed by the initial data
\be\label{rel1-3} S_{a}(t_0,x^{\mu})=S_{a0}(x^{\mu}),\qquad
S_{e}(t_0,x^{\mu})=S_{e0}(x^{\mu})\ee to formulate the standard
Cauchy problem. Our formalism is not equivalent to the conventional
one, because a principal solution of the Hamilton-Jacobi equation
(\ref{rel1-1}) does not satisfy the corresponding part of the
initial condition (\ref{rel1-3}) in general case. However, one can
prove, that our modified dynamics coincides identically with the
standard classical theory in the special case of the
delta-functional probability density \cite{r-1}.

In solving the problem (\ref{rel1-1})--(\ref{rel1-3}), one
constructs the action $S_a$, and calculates the entropy function
$S_e$. After that, it is necessary to introduce the potential
\be\label{rel1-4} S=S_e+\beta \, E;\ee its minimum in respect to
$x^{\mu}$ at the given (but arbitrary) values of the variables $t$
and $\beta$ defines the classical world sheet
$x^{\mu}=x^{\mu}(t,\,\beta)$ of the theory. Here $
E=-{1}/{m}\,\,g_0^{\mu\nu}S_{a,\,\mu}S_{a,\,\nu}$ in the theory
under discussion, and $\beta=1/kT$, where $k$ is the Bolzman
constant and $T$ is the absolute temperature. The statement is that
the probability density $\rho$ is related to this potential as
$\rho=\exp{(- S)}/Z$, where $Z=\int d^4 x \exp{(- S)}$. It is clear,
that any minimum of $S$ corresponds to the maximum of $\rho$. Thus,
we define the classical world sheet for the particle as an extremal
two-dimensional manifold in the space-time, which maximizes locally
its probability density distribution. To extract a classical
trajectory from this world sheet, one must fix the temperature
curve, i.e. to define the corresponding thermodynamic regime
$\beta=\beta (t)$.

Then, to derive dynamical equations of the theory, which define its
classical trajectories, one must calculate first derivatives of the
necessary extremum condition $S_{,\mu}=0$ in respect to the
parameters $t$ and $\beta$. After that, by the help of extracting of
the quantities $x^{\mu}_{,t}$ and $x^{\mu}_{,\beta}$ from the
relations obtained, one must write down the total $t$-derivatives of
the trajectory space-time coordinates
$\dot{x}^{\mu}=x^{\mu}_{,t}+\dot{\beta}x^{\mu}_{,\beta}$, where
$\dot{\beta}=d\beta/dt$. The result reads: \be\label{rel1-5}
\dot{x}^{\mu}=\frac{2}{m}\left [ -g_0^{\mu\nu}p_{\nu}+\left (
A_2^{-1}\right )^{\mu\nu}\Box S_{a,\,\nu}+\dot\beta \left (
A_2^{-1}A_1\right )^{\mu}_{\,\,\nu}g_0^{\nu\lambda}p_{\lambda}
\right ],\ee where $p_{\mu}=S_{a,\,\mu}$,
$A_{1\,\mu\nu}=S_{a,\,\mu\nu}$, $A_{2\,\mu\nu}=S_{,\,\mu\nu}$, and
all the quantities arisen must be calculated on the classical
trajectory under consideration. Note, that in Eq. (\ref{rel1-5}), we
suppose a non-degenerated character of the matrix $A_{2\,\mu\nu}$.
Moreover, below we mean the positive definiteness for this matrix to
guarantee a maximal value of the probability density on the
classical trajectory. To complete the modified Hamilton scheme, it
is necessary to add the dynamical momentum equation
\be\label{rel1-6} \dot{p}_{\mu}=\frac{2}{m}\left [ \left (
A_1A_2^{-1}\right )_{\mu}^{\,\,\nu}\Box S_{a,\,\nu}+\dot\beta \left
( A_1A_2^{-1}A_1\right )_{\mu\nu}g_0^{\nu\lambda}p_{\lambda} \right
]\ee to the coordinate one (5). Our goal is to study the system
(\ref{rel1-5})--(\ref{rel1-6}), and to clarify a role of the
quantities $A_{1\,\mu\nu}$ and $A_{2\,\mu\nu}$ in the generalized
classical dynamics under consideration. We would like to show, that
they generate a curvature in the originally flat space-time of the
theory, making its effective metric actually nontrivial.

First of all, in the case of these quantities absence, one can
integrate the system (\ref{rel1-5})--(\ref{rel1-6}) without any
problem. In fact, to simplify this system, it is enough to put
$(A_{2}^{-1})^{\mu\nu}=0$ only, i.e., to take the probability
density in the form of delta-function with its center on the
classical trajectory of the particle. The resulting world line is
non-curved; also one concludes, that $x^0\sim t$ on this line and,
moreover, that $t$ is proportional to self time of the particle.
However, from Eqs. (\ref{rel1-5})--(\ref{rel1-6}) it follows, that
in the presence of the nontrivial quantities $A_{1\,\mu\nu}$ and
$A_{2\,\mu\nu}$, the classical trajectories will be curved. Note,
that this effect takes place for the free (i.e., for the
potential-free) system, which trajectories are defined by some
extremal principle (by maximum of the probability density, or by
minimum of the potential $S$).

A very similar theoretical construction is related to the standard
description of dynamics of the point mass in the General Relativity
\cite{gr}. In this theory, any world line of the particle must
minimize its classical action, which is proportional to geometrical
length of the trajectory. In the situation with non-trivial metric
field, one obtains curved world lines of the point mass. In the
canonical formalism, the Hamiltonian function for the such particle,
which propagates in the background metric $g_{\mu\nu}$, reads:
\be\label{rel1-7} H=-\frac{1}{m}g^{\mu\nu}p_{\mu}p_{\nu}.\ee The
corresponding dynamical equations are: \be\label{rel1-8}
\dot{x}^{\mu}=-\frac{2}{m}g^{\mu\nu}p_{\nu},\qquad
\dot{p}_{\mu}=\frac{1}{m}g^{\nu\lambda}_{\,\,\,\,\,\,\,\,\,
,\mu}\,p_{\nu}\,p_{\lambda},\ee where it was taken into account,
that $g_{\mu\nu}=g_{\mu\nu}(x^{\lambda})$. Note, that we use the
same notation for the coordinates, momentums, and the dot for
derivative in respect to the same dynamical parameter $t$ for this
system, as for the original one, in view of success of their
following identification.

It is clear, that one can wait coinciding of the our dynamics with
the General Relativity case only in the `near to hard'\, variant of
the first one. Actually, the General Relativity deals with standard
mathematical curves, whereas in the soft variant of the dynamics
under consideration a classical trajectory describes the most
probable world line of the particle. In the complete `soft
picture',\, one has a `probability stream',\, or some `trajectory
tube'\, at least, which cannot be compared with one-dimensional
geodesic line in the curved space-time. Thus, one must put
$(A_{2}^{-1})^{\mu\nu}\sim\epsilon$, where $\epsilon\rightarrow 0,
\, \epsilon\neq 0$, to make the `trajectory tube'\, infinitesimally
thin. In doing so, one obtains, for the second differential of $S$
with the center on the classical trajectory, that $d^2 S\sim
1/\epsilon$, and the exponential of the probability density becomes
of the Gaussian type approximately. The infinitesimal parameter
$\epsilon$ detects a `resonance quality'\, of the probability
density; it gives the delta-functional ideal detection for the point
particle at the limit of $\epsilon=0$. In fact, we are interested in
a study of the `near to hard'\, variant of the `soft'\, classical
dynamics, which describes correctly $\sim\epsilon$ perturbations
under the non-excited `hard'\, theory case.

Thus, let us represent the coordinates and moments of the point mass
into the following linear forms in respect to the parameter
$\epsilon$: \be\label{rel1-10} x^{\mu}=\sum_{k=0}^1
x^{\mu}_{(k)},\qquad p_{\mu}=\sum_{k=0}^1 p_{{(k)}\,\mu},\ee where
terms with the index $`k'$ mean as proportional to $\epsilon^k$.
Then, it is possible to show, that the quantity
$A_{1\,\mu\nu}\sim\epsilon^0$. This statement is supported by the
fact, that the all  matrix quantities $(A_{2}^{-1})^{\mu\nu}, \,
(A_2^{-1}A_1)^{\mu}_{\,\,\,\nu}$ and $(A_1A_2^{-1}A_1)_{\mu\nu}$
have a common correlation nature and must be of identic excitation
power ($\sim\epsilon^1$,\, like the first of them). Actually, it is
not difficult to prove, that these quantities describe leading
$\sim\epsilon$ terms of the correlations $x^{\mu}\circ x^{\nu}$,
$x^{\mu}\circ p_{\nu}$ and $p_{\mu}\circ p_{\nu}$ of the variables
$x^{\mu}$ and $x^{\nu}$, $x^{\mu}$ and $p_{\nu}$, and $p_{\mu}$ and
$p_{\nu}$, respectively. Here, the correlation $A\circ B$ of the
quantities $A$ and $B$ is defined as $A\circ
B=\overline{AB}+\overline{BA}-2\overline{A}\,\,\overline{B}$. In the
classical variant under consideration, all the average values must
be calculated using the probability density $\rho$ introduced above,
and with taking into account of the relation
$p_{\mu}=S_{a,\,\,\mu}$.

Now let us substitute the decompositions (\ref{rel1-10}) into the
equations (\ref{rel1-5}) and (\ref{rel1-6}). For the
$\sim\epsilon^0$ terms, one obtains the following equations:
\be\label{rel1-12} \dot
{x}^{\mu}_{(0)}\,\,=-\frac{2}{m}p_{(0)}^{\mu},\qquad \dot
{p}_{{(0)}\,\mu}=0,\ee where
$p_{(0)}^{\mu}=g_{(0)}^{\mu\nu}p_{{(0)}\,\nu}$. Solving them, one
concludes, that ${p}_{(0)}^{\mu}={\rm const.}$, and in the
$\epsilon^0$-approximation world lines of the theory are non-curved
actually. Then, for the $\sim \epsilon$ terms, the dynamical
equations read: \be\label{rel1-13}\dot
{x}^{\mu}_{(1)}\,\,&=&\frac{2}{m}\left [
-g_{(0)}^{\mu\nu}p_{{(1)}\,\nu}+\left ( A_2^{-1}\right
)^{\mu\nu}\Box S_{a,\,\nu}+\dot\beta \left ( A_2^{-1}A_1\right
)^{\mu}_{\,\,\nu}p_{(0)}^{\nu}
\right ],\nonumber\\
\dot {p}_{{(1)}\,\mu}&=&\frac{2}{m}\left [ \left ( A_1A_2^{-1}\right
)_{\mu}^{\,\,\nu}\Box S_{a,\,\nu}+\dot\beta \left (
A_1A_2^{-1}A_1\right )_{\mu\nu}p_{(0)}^{\nu} \right ].\ee It is
seen, that their nontrivial dynamics is impossible without
non-vanishing correlations of the canonical variables of the theory.

Our next step is to analyze the system (\ref{rel1-8}) in the same
perturbation manner. First of all, let us represent the metric field
$g_{\mu\nu}$ in the form of Eq. (\ref{rel1-10}): \be\label{rel1-14}
g_{\mu\nu}=g_{(0)\,\mu\nu}+g_{(1)\,\mu\nu}.\ee In doing so, we
suppose, that $g_{(1)\,\mu\nu}\sim \epsilon$, i.e. that the
background metric of the system under consideration has a small
excitation under the flat Minkowsky part. After the substitution of
Eq. (\ref{rel1-10}), (\ref{rel1-14}) into Eq. (\ref{rel1-8}), one
obtains exactly the system (\ref{rel1-12}) for the $\sim\epsilon^0$
terms. Then, for the $\sim \epsilon$ terms, the dynamical equations
read: \be\label{rel1-15} \dot {x}^{\mu}_{(1)}\,\,&=&
\frac{2}{m}\left [ -g_{(0)}^{\mu\nu}p_{{(1)}\,\nu}
+g_{(0)}^{\mu\lambda}p_{(0)}^{\sigma}
g_{(1)\,\lambda\sigma}\right ],\nonumber\\
\dot
{p}_{(1)\,\mu}&=&-\frac{1}{m}g_{(1)\,\nu\lambda,\,\mu}p_{(0)}^{\nu}p_{(0)}^{\lambda}.\ee
Our goal is to identify Eqs. (\ref{rel1-13}) and (\ref{rel1-15}) --
i.e., to express the metric excitation $g_{(1)\,\mu\nu}$ and its
first derivatives in terms of correlations and other characteristics
of the original classical system.

To do this, it is convenient to choose a special coordinate system
with $p_{(0)}^{\mu}=p_{(0)}^0\delta_0^{\mu}$, where $p_{(0)}^0=-mc$
for the `near to hard'\, classical scheme under consideration and in
view of the metrics signature taken. Thus, we perform our
identification in the initial rest system of the relativistic
particle. The result reads: \be\label{rel1-16} g_{(1)\,\,
0\lambda}&=&g_{(0)\,\lambda\mu}I^{\mu}, \qquad g_{(1)\,\,
00,\,\lambda}=\frac{2}{mc}A_{1\,\lambda\mu}I^{\mu},\ee where
\be\label{rel1-17} I^{\mu}=\left ( A_2^{-1}\right )^{\mu\nu}\left
(-\frac{1}{mc}\Box S_{a,\,\nu}+\dot\beta A_{1\,\,\nu 0}\right ).\ee
Now, using Eq. (\ref{rel1-16}) and the standard non-relativistic
relation between $g_{(1)\,\, 00}$ and the weak gravitational
potential, one can write down an explicit form of the latter one or
of the corresponding gravitational force. However, here we would
like to clarify another and much more principal question, which is
related to unclear consistency of Eqs. (\ref{rel1-16}) and
(\ref{rel1-17}). Namely, using these two relations, one can
calculate the quantity $\dot g_{(1)\,\,00}$ in two different ways.
Actually, one can calculate it, using straightforward total
differentiation of the right side of the first relation from Eq.
(\ref{rel1-16}) in respect to the thermodynamical time $t$. The
alternative way is to use the second relation from Eq.
(\ref{rel1-16}), and the standard chain rule $\dot
g_{(1)\,\,00}=g_{(1)\,\, 00,\,\mu}\,\dot x^{\mu}\approx g_{(1)\,\,
00,\,0}\dot x^{0}_0$ (the last approximation becomes evident after
taking into account only the $\sim\epsilon$ terms). Finally, one
obtains the following consistency condition for the identification
scheme established: \be\label{rel1-18}\dot
I^{0}=\frac{4}{m}A_{1\,\,0\mu}I^{\mu}.\ee

To clarify a general status of this relation, one must note, that
its left side can contain the therm $\sim\ddot\beta$, whereas its
right side is free of the such type of quantities. Thus, Eq.
(\ref{rel1-18}) means the differential equation of the second (or of
the first -- see below) order, which must be satisfied by the
temperature curve $\beta=\beta(t)$ to complete our identification of
the generalized relativistic dynamics under consideration with the
propagation of the point particle along geodesic line in the curved
background space-time. The explicit form of this relation reads:
\be\label{rel1-19} \dot I^{0}&=&\left ( A_{2}^{-1}\right
)^{0\nu}\left \{ \ddot\beta S_{a,\,\,00\nu}+ \frac{1}{m}\left [
\left (-\frac{1}{mc}\Box+\dot\beta\partial_0\right
)\partial_\nu\left (
g_{(0)}^{\lambda\sigma}S_{a,\,\,\lambda}S_{a,\,\,\sigma}+2mcS_{a,\,\,0}\right
)\right ]\right \}\nonumber\\&-&\frac{2}{m}\left
(g_{(0)}A_{1}+A_2^{-1}A_1g_{(0)}A_2\right
)^{0}_{\,\,\,\nu}I^{\nu},\ee where only the leading $\sim\epsilon$
terms had been written down in the its right side. Of course, the
all quantities there must be calculated on the classical trajectory
under consideration. Finally, one must substitute $\dot I^{0}$ from
Eq. (\ref{rel1-19}) to Eq. (\ref{rel1-18}), and to explore Eq.
(\ref{rel1-17}) for detecting of the temperature regime which must
be taken in the identification scheme constructed. Namely, the
second relation from Eq. (\ref{rel1-16}), together with the
definition (\ref{rel1-17}) of the quantity $I^{\mu}$, and the
thermodynamic regime established define the effective gravitational
force completely. This force acts on the relativistic point mass
exactly as it takes place in the standard General Relativity. This
means a possibility to interpret the gravitational force not only in
the geometric, but also in the correlations related framework.

Here we would like to make several notes. The first one is related
to the fact, that there is not one-to-one correspondence between the
standard motion of a point mass in the weakly curved metric field,
and the propagation of the same object in regime of the small
correlations in the generalized dynamics developed. Actually, it is
clear, that the general temperature regime leads to the
non-geometrizing effects in this dynamics. Thus, these two theories
have the common part of the `weak regions',\, but do not coincide
identically. However, one can say about some new `non-Einstein'\,
realization of the gravity theory in framework of the our approach
for the arbitrary thermodynamic regime of the system evolution. This
new approach is preferable in comparison with the standard
geometrical ones, because it does not needs in any additional
axiomatic inputs instead of the ones arising naturally from the
quantum theory. We would like to note, that our `minimal
modification'\, of the quantum theory allows one to incorporate
arrow of time into the gravity dynamics without lose of its
fundamental nature. Actually, introducing of the arrow of time into
the theory dynamics is guaranteed by non-Hermitian Hamiltonian
structure, see \cite{q-1}--\cite{q-c}. This seems especially
promising for the black hole \cite{bh-f}--\cite{bh-l} and
cosmological \cite{cosm} applications, where irreversibility of
evolution must be guaranteed without violation of the general
theoretical scheme.

Our special interest is related to extension of the formalism
developed to the string theory case \cite{math}--\cite{kir}. For
example, in pure gravitational context, it is much more natural to
deal with a string than with a point particle. The reason is the
following: using identification scheme of the developed type, one
can obtain the metric coefficients on the classical string
two-dimensional world sheet, instead of the on-dimensional world
line of the particle. From this point of view, to reach the maximal
analogy with the General Relativity, one must develop a relativistic
theory of the four-dimensional `fluid',\, which complete the whole
physical space-time. One can start from the corresponding quadratic
Polyakov-type Hamiltonian, which generalizes the one (\ref{rel1-7}),
used for the zero-dimensional manifold in this paper. In this
situation, the classical correlations generate effective metric
excitations under the total four-dimensional space-time of the
theory. The corresponding classical gravitation can be studied in
the precise General Relativity manner.

Finally, we would like to stress, that the our approach is really
universal. The only necessary element for its realization is related
to construction of the Poincare-invariant Hamiltonian $H$. We take
it from the standard theory, and after that it is necessary only to
develop all well-defined quantum and corresponding classical
constructions. They differ from the standard ones in one really
fundamental part: the new classical dynamics includes the full set
of correlations for the canonical variables. These quantities give
rise to arrow of time in the case of the non-Hermitian Hamiltonian
structure, as well as to generation of the effective gravity in the
free system case.


\vskip 1cm

\noindent {\large \bf Acknowledgements}

\vskip 3mm \noindent We would like to thank prof. B.S. Ishkhanov for
many discussions and private talks which were really useful for us
during this work preparation. One of the authors (O.V.K) was
supported by grant ${\rm MD \,\, 3623.\, 2006.\, 2}$.

\end{document}